\documentclass[10pt,a4paper]{article}
\usepackage{jheppub_kim}
\usepackage{pdflscape}
\usepackage{amsmath}
\usepackage{amssymb}
\usepackage{dcolumn}
\usepackage{bm}
\usepackage{color}
\usepackage{epsfig}
\usepackage{amsfonts}
\usepackage{graphicx}
\usepackage{subfigure}
\usepackage{dcolumn}

\begin{document}
\title{A new type of $f(\mathcal{T})$ gravity from Barrow entropy}

\author[a]{Tuhina Ghorui,}
\author[b]{Prabir Rudra,}
\author[c]{Farook Rahaman}

\affiliation[a]{Department of Mathematics, Jadavpur University,
Kolkata-700 032, India.}

\affiliation[b] {Department of Mathematics, Asutosh College,
Kolkata-700 026, India.}

\affiliation[c] {Department of Mathematics, Jadavpur University,
Kolkata-700 032, India.}

\emailAdd{tuhinaghorui.math@gmail.com}
\emailAdd{prudra.math@gmail.com}
\emailAdd{farookrahaman@gmail.com}

\abstract{In this work, two originally separate adjustments for the Friedmann equations are concurrently considered. Firstly, the fractal structure of the black hole horizon region is imposed by the Barrow entropy. The second adjustment is the $f(\mathcal{T})$ gravity, which is based on a teleparallel framework generalization of the Einstein-Hilbert action, where $\mathcal{T}$ is the scalar torsion. This can be considered as a Barrow entropy modification of the $f(\mathcal{T})$ gravity thus yielding a new model. Gravity thermodynamics hypothesis principles are used to integrate these two models under a unified framework. We derive the modified Friedmann equation and note the corrections obtained. To understand the implications of such dual modifications, an application is analyzed and a particular $f(\mathcal{T})$ toy model is chosen for the purpose. The equation of the state parameter of the resulting model, the dimensionless density parameters of matter and dark energy, and the deceleration parameter are explored to check the viability of the new model. A discussion of the dynamic evolution of the universe follows from these results. It is seen that the results comply with the observations. The newly developed model is promising and demands further study.}


\maketitle

\section{Introduction}
One of the primary forces in the universe is gravity. Nevertheless, no comprehensive explanation of this connection exists covering all energy levels. The most effective theory of gravitation from a classical perspective is general relativity (GR), which agrees extremely well with numerous observational tests \cite{cm1}. While discussions and the construction of quantum versions have been ongoing, there is still no comprehensive theory of quantum gravity. At low energies, some efficient research has been done on some theories \cite{jfd,jfd1,jfd2,jfd3,cpb}. Black hole thermodynamics is a crucial field of study that provides basic knowledge about the quantum nature of gravity. A black hole radiates at a certain temperature, and its entropy is related to the area of its event horizon, as demonstrated by early research by Hawking and Bekenstein\cite{jdb1, swh, JDB}. This suggests that black hole thermodynamics may be explored within the framework of quantum gravity to retrieve crucial information. In addition, since the initial commentary concerning potential connections between thermodynamics and general relativity, suggestions have surfaced regarding how to derive formulas that more accurately characterize the various stages of the universe's expansion. The holographic principle  \cite{RB} is one of the concepts that examines a system's behavior by examining the attributes of its exterior only. Such theories seek to modify the entropy calculation procedure \cite{swh, SWH}. Other conjectures, such as the gravity-thermodynamics conjecture, which examines the connection between thermodynamics and Einstein equations, have been examined in light of these findings.

By utilizing the connection between the entropy and the black hole's event horizon region\cite{tp2, tp1}, the gravity-thermodynamics conjecture is a formalism that makes it possible to derive Einstein's equations from a thermodynamic perspective\cite{RGC, TJ}. The Friedmann equations can be derived by applying the first law of thermodynamics to the apparent horizon. This process can be used in general relativity and several modified gravity theories. Nonetheless, the entropy relation typically alters when the research is expanded to a revised theory \cite{ABG, RM8, RMW}. Various generalized statistical mechanics have been proposed to explore cosmic evolution and gravitational phenomena in general since the Bekenstein-Hawking entropy-area law is a non-extensive measure. This has led to the development of other models, including Renyi entropy \cite{AR1} and Tsallis entropy \cite{CT, LJ}. A new theory called Barrow entropy \cite{jb} has recently been given some thought. This paper will examine Barrow entropy and its implications for the evolution of the universe.

Barrow was motivated to create his model by the COVID-19 virus's geometry structure \cite{jb}. According to this hypothesis, the black-hole structure may have complex fractal patterns due to quantum-gravitational forces. The horizon surface of this system is fractal in nature. Using this structure the new black hole entropy relation is defined as $S\sim A^{f(\Delta)}$, where $f(\Delta)=1+\frac{\Delta}{2}$. Quantifying the quantum-gravitational deformation is done through the $\Delta$ parameter. Within this framework, certain applications have been explored recently. For instance, research on Baryon asymmetry \cite{gl} and inflation caused by Barrow Holographic dark energy has been examined from a new perspective in the early and late phases of the cosmos \cite{sp}. Barrow holographic dark energy has been developed\cite{ens}, and generalized entropy has been examined \cite{sst}. Barrow entropy has been studied in detail in \cite{es}.

Several observational data points reveal the accelerated expansion of the current universe \cite{es, sc, as, ep, mn, cb} which provides impetus for considering an alternative gravity theory to general relativity. Over time, a multitude of proposals for different forms of modified gravity have been implemented to tackle a broad spectrum of physical problems \cite{ft1, ft2, ft3, ft4, ft5, ft6, bib16, bib19}. It is significant to remember that the basis of GR is Riemannian geometry, which is based on curvature, derived from the metric. This happened mainly because, at the time GR was proposed, curvature was the only properly developed geometric tool available \cite{ft7}. Over the following decades, other geometric approaches have been properly developed, such as torsional models contained in teleparallel gravity (TG) \cite{ft8, ft9, ft10, ft11}. The basis of TG is the exchange of the curvature-based Levi-Civita connection with the torsion-based teleparallel connection, which has become incredibly popular in recent years. Considering new physics at the level of gravitational theory, this picture may provide a more rational approach. The teleparallel link is curvature-free, satisfies metricity and serves as the cornerstone of a novel gravitational framework \cite{ft12}. However, a specific set of gravitational contractions leads to the teleparallel equivalent of general relativity (TEGR), which is dynamically equal to GR at the level of the classical field equations (see \cite{ft13} for a discussion). As such, both GR and TEGR have similar predictions for astrophysical and cosmic phenomenology; nevertheless, differences may arise when considering IR completions \cite{ft14}. The action in the torsion scalar $\mathcal{T}$ is linear in this instance. Changes to TG through the TEGR action can now be investigated using the same logic that applies to conventional modified gravity in the GR regime. One popular technique for generalizing TEGR is $f(\mathcal{T})$ teleparallel gravity, which produces generically second-order equations of motion for all spacetimes based on an arbitrary function of the torsion scalar \cite{ft15, ft16, ft17, ft18, ft19, ft20, ft21}. other developments of this have taken numerous forms, such as the use of a scalar field \cite{ft22, ft23, ft24} and other scalar modifications \cite{ft25, ft26}, like regular curvature-based modified gravity. 

In this work, we focus on the study of $f(\mathcal{T})$ gravity in conjunction with Barrow entropy, utilizing the concepts of the gravity-thermodynamics conjecture to describe the universe's evolution. The amalgamation of a torsion-based gravity with the quantum gravitational effects and fractal structure of Barrow entropy is expected to produce an interesting cosmological model, with far-reaching implications. The structure of this paper is as follows: The Barrow entropy and its associated intricacies are explained in section 2. Section 3 presents the basic equations of $f(\mathcal{T})$ gravity and the first law of thermodynamics. Barrow entropy is incorporated in the framework of $f(\mathcal{T})$ theory in section 4. Section 5 examines an application of the work with a toy model. Finally, the paper ends with a conclusion in section 6

\section{Barrow Entropy}
In this section, we use the gravity-thermodynamics conjecture \cite{mr} to derive the Friedmann equations. Beginning with the first law of thermodynamics, two distinct scenarios are examined. The Bekenstein-Hawking entropy is employed in the first scenario, while the Barrow entropy is utilized in the second scenario. The first law of thermodynamics is given by,
\begin{equation}
T dS=dE-WdV,
\end{equation}
where T, S, E, and V are the temperature, entropy, energy, and volume, respectively, and $W=\frac{(\rho-p)}{2}$ is the work, where $\rho$ and p stand for energy density and pressure, respectively. It is necessary to understand temperature and entropy to apply the first law of thermodynamics. The Bekenstein-Hawking entropy, or $S=\frac{A}{4G}$, is the commonly used method. Here, A presents the black hole horizon's area ($A=4\pi r^{2}$), and G is the gravitational constant \cite{sb}. The horizon temperature, $T=-\frac{1}{2\pi r}\left(1-\frac{\dot r}{2 H r}\right)$, is selected for the temperature \cite{RGC, ad4}. However, since we are working with the flat case of the FRW metric, which implies $k = 0,$ we have $Hr = 1.$ Normally, r the apparent horizon, would be $ r=\frac{1}{\sqrt{H^{2}+\frac{k}{a^{2}}}}$. As a result, the temperature can be expressed more simply as $ T=-\frac{1}{2\pi r}\left(1-\frac{\dot r}{2}\right) $ for flat FRW spacetime. The first law of thermodynamics then turns into
\begin{equation} \label{a}
-\frac{1}{2\pi r}\left(1-\frac{\dot r}{2}\right)\frac{dA}{4G}=dE-\frac{\rho-p}{2}dV
\end{equation}
Equation (\ref{a}) can be expressed by applying the matter conservation equation, (which is $\dot{\rho}+3H(\rho+p)=0$) as.
\begin{equation}
-\frac{1}{2\pi r}\left(1-\frac{\dot r}{2}\right)\frac{dA}{4G}=-A\left(\rho+p\right)dt+\frac{\rho+p}{2}dV
\end{equation}
where it is stated that $dE = Vd\rho+ \rho dV = -A(\rho+p)dt+\rho dV$. Employing $dV = Adr = A\dot{r}dt$ results in
\begin{equation}\label{l1}
\frac{\dot r}{r^{2}}=4 \pi G\left(\rho+p\right)
\end{equation}
when $\frac{\dot{r}}{r^{2}}=-\dot{H}$ occurs in a flat universe. Therefore equation (\ref{l1}) becomes
\begin{equation}
\dot{H}=-4 \pi G\left(\rho+p\right)
\end{equation}
This is the Friedmann equation. When the above equation is integrated in time and applied to the conservation relation, we have
\begin{equation}
H^{2}=\frac{8\pi G}{3}\rho + C
\end{equation}
where $C$ acts as the cosmological constant.

It is significant to remember that there is another method to get the same Friedmann equation when taking into account a fixed boundary $(\dot{r}= 0)$ with an energy flux through it \cite{RGC}. In this instance, the first law of thermodynamics is applied, with $dE=TdS$, and the temperature is defined as $T=\frac{1}{2\pi r_{A}}$ for a static system.

There are some proposals for obtaining new equations of motion. One of them is to consider
modifications in the black hole area, as proposed by Barrow \cite{jb}. In this model, the area takes into
account a possible fractal structure from the surface roughness, which depends on a parameter $\Delta$ representing its intricacy. This leads to a new form of entropy, which is given as

\begin{equation}
S=\left(\frac{A}{4G}\right)^{1+\frac{\Delta}{2}}
\end{equation}
Using the same method as before, the equation of motion or the modified Friedmann equation can be derived from the Barrow entropy. Let's look at the entropy defined as $S=\frac{f(A)}{4G}$ in a more general sense. In our instance, the Barrow definition is matched by selecting $f(A) = A^{1+\frac{\Delta}{2}}/(4G)^{\frac{\Delta}{2}}$. Beginning with the thermodynamics first law, we get
\begin{equation}\label{l2}
-\frac{1}{2\pi r}\left(1-\frac{\dot r}{2}\right)\frac{df(A)}{4G}=dE-\frac{\rho-p}{2}dV
\end{equation}
As $df(A)=f_{A}dA$, equation (\ref{l2}) takes the form:
\begin{equation}
-f_{A}\frac{1}{2\pi r}\left(1-\frac{\dot r}{2}\right)\frac{dA}{4G}=dE-\frac{\rho-p}{2}dV
\end{equation}
It should be mentioned that this equation and Eqn.(\ref{a}) are comparable. Then, using the same process as previously, it is determined that
\begin{equation} \label{b}
f_{A}\dot{H}=-4 \pi G\left(\rho+p\right)
\end{equation}
Barrow entropy has led to a modification in the Friedmann equation. Equation (\ref{b}) is used in the energy conservation relation to derive the second Friedmann equation as
\begin{equation}
\dot{\rho}=\frac{3}{4 \pi G} f_{A} H \dot{H}
\end{equation}
Integrating we get,
\begin{equation} \label{c}
\frac{2}{2-\Delta} f_{A}H^{2}=\frac{8 \pi G}{3}\rho+\frac{\Lambda}{3}
\end{equation}
where the integration constant $\Lambda=3\frac{2}{2-\Lambda}f_{A_{0}}H_{0}^{2}-8\pi G\rho_{0}$ is used. Here $\rho_{0}$ and $H_{0}$ stand for the present values of fluid's energy density and the Hubble parameter, respectively. The horizon area is also determined from these conditions using the formula $A_{0}=4\pi H_{0}^{-2}$. It is significant to remember that the function $f(A)$ is the only thing that this version of equation (\ref{c}) depends on.

The structure of these equations is identical to that of the standard Friedmann equations, demonstrating the existence of dark energy derived from the new entropy. Equation (\ref{b}) is provided as
\begin{equation}\label{e}
\dot{H}=-4 \pi G\left(\rho+p-\frac{\dot{H}}{4 \pi G}\left(1-f_{A}\right)\right)
\end{equation}
and equation (\ref{c}) turn into
\begin{equation} \label{d}
H^{2}=\frac{8 \pi G}{3}\left(\rho+\frac{\Lambda}{8 \pi G}+\frac{3H^{2}}{8 \pi G}\left(1-\frac{2}{2-\Delta}f_{A}\right)\right)
\end{equation}
Because of this, the elements of dark energy become
\begin{equation} \label{f}
\rho_{DE}=\frac{1}{8 \pi G}\left(\Lambda(\Delta)+3H^{2}\left(1-\frac{2}{2-\Delta}f_{A}\right)\right)
\end{equation}
\begin{equation} \label{g}
p_{DE}=-\frac{1}{8 \pi G}\left(\Lambda(\Delta)+2\dot{H}(1-f_{A})+3H^{2}\left(1-\frac{2}{2-\Delta}f_{A}\right)\right)
\end{equation}
$\rho_{de}$ and $p_{de}$ are Barrow's holographic dark energy density and pressure respectively \cite{jn}. It is important to note that $f(A)=A^{1+\frac{\Delta}{2}}/{(4G)^{\frac{\Delta}{2}}}$ is specifically taken into consideration for Eq. (\ref{d}) as it is for $\rho_{de}$ and $p_{de}$.

It is significant to note that, while Eqs. (\ref{b})–(\ref{c}) suffice to study the universe's dynamics, the conservation relations $\dot{\rho}_{m}+3H\left(\rho_{m}+p_{m}\right)$ and $\dot{\rho}_{de}+ 3H(\rho_{de} + p_{de}) = 0$ require the manipulations of Eqs.(\ref{e})–(\ref{d}) that result in the definitions of Eqs. (\ref{f})–(\ref{g}). Consequently, a similar result is shown in references \cite{st,fe,su,yz,sv,tp}, further demonstrating the lack of interaction between these two sectors.

A static system with an energy flux crossing its boundary can be regarded, as in \cite{jn}, just like in the Bekenstein-Hawking case. Under the assumption that $T =\frac{1}{2\pi r_{A}}$ and that $-dE = TdS$, the equations of motion eq. (\ref{d}) and dark energy components eqns. (\ref{f}) and (\ref{g}) can be obtained.
The $f(\mathcal{T})$ gravity is examined in the gravity-thermodynamics conjecture in the following section.

\section{$f(\mathcal{T})$ gravity and first law of thermodynamics}
In this section, we discuss the basic concepts $f(\mathcal{T})$ gravity. We go on to discuss the pressure and effective energy density arising from the torsion-based modified gravity framework. Then the first law of thermodynamics is examined using these quantities. Subsequently an analysis of the black hole entropy correction resulting from $f(\mathcal{T})$ gravity is examined further in this discussion. This model generalizes the Einstein-Hilbert action by substituting a function of the torsion scalar $f(\mathcal{T})$ for the torsion $\mathcal{T}$, in TEGR i.e.,
 
\begin{equation}
S=\int d^{4}x   e \left [\frac{f(\mathcal{T})}{16\pi G}+I_{m}\right]
\end{equation}
it is known that the variational principle is the source of field equations. When the action is varied about the metric, the field equations are generated as
\begin{equation}
\frac{1}{e}\partial_{\mu}\left(e e^{\rho}_{A}S_{\rho}^{\mu\nu}\right)\left [1+f_{\mathcal{T}}\right]-e_{A}^{\lambda}\mathcal{T}^{\rho}_{\mu\lambda} S_{\rho}^{\nu\mu}\left[1+f_{\mathcal{T}}\right] + e^{\rho}_{A}S_{\rho}^{\mu\nu} \left(\partial_{\mu}\mathcal{T} \right) f_{\mathcal{T}{T} }+ \frac{1}{4}e_{A}^{\nu} \left[\mathcal{T}+ f(\mathcal{T})\right] = 4\pi G e^{\rho}_{A} \mathcal{T}^{(m)^{\nu}}_{\rho} 
\end{equation}
The effective Friedmann equations become
\begin{equation}
 H^{2} =\frac{8\pi G}{3}\rho_{m}-\frac{f(\mathcal{T})}{6}-2f_{\mathcal{T}}H^{2}
 \end{equation}
 
 and
 
 \begin{equation}
 \dot{H}=-\frac{4\pi G\left(\rho_{m}+p_{m}\right)}{1+f_{\mathcal{T}}-12H^{2}f_{\mathcal{TT}}}
 \end{equation}
Where $\kappa^{2}=8 \pi G$. With a flat FRW universe and a perfect fluid as the matter content, the field equations are as follows:
\begin{equation}\label{t1}
H^{2}=\frac{\kappa^{2}}{3f_{\mathcal{T}}}\left(\rho_{m}+\rho_\mathcal{T}\right)
\end{equation}
\begin{equation}\label{t2}
2\dot{H}+3H^{2}=-\frac{\kappa^{2}}{f_{\mathcal{T}}}\left(p_{m}+p_{\mathcal{T}}\right)
\end{equation}
where the effective density and pressure are defined as
\begin{equation}
\rho_{\mathcal{T}}=\frac{1}{2k^{2}}\left(\mathcal{T}f_{\mathcal{T}}-f\right)
\end{equation}
and
\begin{equation}
p_\mathcal{T}=\frac{1}{2\kappa^{2}}\left(f-\mathcal{T}f_\mathcal{T}+4H\dot{\mathcal{T}}f_\mathcal{TT}\right)
\end{equation}
The non-conservation equation is given by
\begin{equation} \label{h}
\dot\rho_{\mathcal{T}}+3H\left(\rho_{\mathcal{T}}+p_{\mathcal{T}}\right)=\frac{3}{\kappa^{2}}H^{2}\dot{\mathcal{T}}f_{\mathcal{TT}}
\end{equation}
Using Eqn.(\ref{t1}) in (\ref{t2}) we get,
\begin{equation}
dH=-\frac{\kappa^{2}}{2f_{\mathcal{T}}}\left(\rho_{m}+p_{m}+\rho_{\mathcal{T}}+p_{\mathcal{T}}\right)
\end{equation}
considering relation $Hr=1$, the last equation become
\begin{equation}\label{l5}
\frac{f_{\mathcal{T}}}{G}dr=A\left(p_{m}+p_{\mathcal{T}}+\rho_{m}+\rho_{\mathcal{T}}\right)
\end{equation} 
The black hole entropy and its relationship to the horizon region are adjusted in this gravitational theory. Here, the entropy is expressed as 
\begin{equation}\label{k}
S=\frac{Af'(\mathcal{T})}{4G}
\end{equation}
The equation using this updated black hole entropy is
\begin{equation}
\frac{1}{2\pi r}dS-\frac{1}{2\pi r}\frac{A}{4G}df'(\mathcal{T})=A\left(p_{m}+p_{\mathcal{T}}+\rho_{m}+\rho_{\mathcal{T}}\right)
\end{equation}
We take the horizon temperature as $T=-\frac{1}{2\pi r}\left(1-\frac{\dot {r}}{2}\right)$ and obtain eqn.(\ref{l5}) as
\begin{equation} \label{i}
Tds-T\frac{A}{4G}df'(\mathcal{T})=-A(\rho+p)ds+\frac{1}{2}A(\rho+p)dr
\end{equation}
where $\rho=\rho_{m}+\rho_{\mathcal{T}}$ and $p=p_{m}+p_{\mathcal{T}}$. The non-conservation relation for $f(\mathcal{T})$ gravity in Eq. (\ref{h}) is added to the matter conservation equation, which is $\dot{\rho}_{m}+ 3H(\rho_{m} + p_{m}) = 0 $, to obtain
\begin{equation}
d\rho=-3H(\rho+p)dt+\frac{3}{{\kappa}^{2}}H^{2} \dot{\mathcal{T}}f_\mathcal{TT}
\end{equation}
As a result, the energy differential represented as $dE=\rho dV + V d\rho$ leads to
\begin{equation} \label{j}
  dE=\rho Adr-A(\rho+p)dt+\frac{1}{2\pi r}\frac{A}{4G}df_\mathcal{T}  
\end{equation}
Equations (\ref{i}) and (\ref{j}) together provide us
\begin{equation}
Tds=dE-WdV+T\left(\frac{4-\dot{r}}{2-\dot{r}}\right)\frac{A}{4G}df_\mathcal{T}
\end{equation}
This is the first thermodynamic law that was developed about gravity. It is crucial to remember that an additional term $ \mathcal{T} \Bar{ds}$ with $\tilde{ds}=-\left(\frac{4-\dot{r}}{2-\dot{r}}\right)\frac{A}{4G} 
df_{\mathcal{T}}$ appears. Several discussions in the literature \cite{ys,tn} make the case that the extra term could have its origins in the internal entropy processed by the system that is out of equilibrium. 
 
Using the Barrow definition as a guide, let's write the modified Friedmann equations found in Eqs. (\ref{t1}) and (\ref{t2}) to derive the first law of thermodynamics without the addition of a term originating from an out-of-equilibrium system.
After that, these equations turn into
\begin{equation}
H^{2}=\frac{\kappa^{2}}{3}\left(\rho_{m}+\rho_{de}\right)
\end{equation}
\begin{equation}
2\dot{H}+3H^{2}=-\kappa^{2}\left(p_{m}+p_{de}\right)
\end{equation}
Where
\begin{equation} \label{l}
\rho_{de}=\rho_{\mathcal{T}}+\frac{3H^{2}}{\kappa^{2}}\left(1-f_{\mathcal{T}}\right)
\end{equation}
\begin{equation} \label{m}
p_{de}=p_{\mathcal{T}}-\frac{1}{\kappa^{2}}\left(2\dot{H}(1-f_{\mathcal{T}})+3H^{2}(1-f_{\mathcal{T}})\right)
\end{equation}
Note that holographic dark energy is described by equations (\ref{l}) and (\ref{m}) because of the $f(\mathcal T)$ theory. The result of these definitions is $\dot{\rho}_{de}+3H\left(\rho_{de}+p_{de}\right)=0$. Applying the previous process, one discovers 
\begin{equation}
    TdS=dE-WdV
\end{equation}.
The standard first law of thermodynamics is thus derived. Here, the entropy correction resulting from the $f(\mathcal T)$ theory (Eq.(\ref{k})), is not applied. In the next section, we add Barrow entropy to $f(\mathcal{T})$ gravity and study the dynamics of the universe described by the new model.

\section{Adding Barrow entropy to $f(\mathcal{T})$ gravity}
Barrow made a well-reasoned and elegant proposal while proposing the new entropy. Nevertheless, some barriers make it impossible to directly apply it to explain the universe's past and present behavior. Among these is the fact that this modification always results in an expansion of the de-sitter type \cite{gj}. An additional issue pertains to the severe limitations placed on the range of values of $\Delta$ when taking into account the properties of the Big Bang \cite{jn, km}. Similarly, the $f(\mathcal{T})$ gravity models also have several issues. These include complex mass values for the scalaron field and general viability conditions like $f_{\mathcal{T}} > 0$ and $f_{\mathcal{TT}} > 0$ to prevent ghost scalar fields. A summary of $f(\mathcal{T})$ gravity and its issues can be found in \cite{ys}.

Here the primary goal is to suggest a method of combining the two models because they each have certain drawbacks. By doing so, we hope to achieve results that are not impacted by these drawbacks. It is possible to see similarities between the definitions provided in Equations (\ref{f}), (\ref{g}), (\ref{l}), and (\ref{m}), as they all refer to modifications that act as dark energy. In light of this, a revised definition of holographic dark energy is taken into consideration, one that includes both the Barrow and $f(\mathcal{T})$ modifications \cite{pa}
\begin{equation} \label{n}
\rho_{de}=\frac{1}{\kappa^{2}}\left(\rho_{\mathcal{T}}+3H^{2}\left(1-\frac{2}{2-\Delta}f_{A}f_{\mathcal{T}}\right)\right)
\end{equation}
\begin{equation} \label{o}
p_{de}=\frac{1}{\kappa^{2}}\left(p_{\mathcal{T}}-2\dot{H}(1-f_{A}f_{\mathcal{T}})-3H^{2}(1-\frac{2}{2-\Delta}f_{A}f_{\mathcal{T}})\right)
\end{equation}
where the relation $\dot{\rho}_{de}+3H(\rho_{de}+ p_{de})= 0$ is satisfied by equations (\ref{n}) and (\ref{o}). Let's begin with the first law to get the modified Friedmann equations.
\begin{equation}
\mathcal{T}=dE-WdV,
\end{equation}
Utilizing $dE=\rho dV+Vd\rho$ and $W=(\rho+p)/2$, we get
\begin{equation}
Tds=Vd\rho+\frac{\rho+p}{2}dV
\end{equation}
Using the energy conservation relation and assuming $\rho=\rho_{m}+\rho_{de}$ and $p=p_{m}+p_{de}$, the final equation becomes
\begin{equation}
Tds=-\left(1-\frac{\dot{r}}{2}\right)A(\rho+p)dt.
\end{equation}
By taking the entropy $S=A/4G$ and the horizon temperature $T=-\frac{1}{2\pi r}\left(1-\frac{\dot{r}}{2}\right)$, we get
\begin{equation}
-\frac{1}{2\pi r}\left(1-\frac{\dot{r}}{2}\right)\frac{dA}{4G}=-\left(1-\frac{\dot{r}}{2}\right)A(\rho+p)dt.
\end{equation}
Following a few steps, it is expressed as
\begin{equation}
\frac{\dot{r}}{r^{2}}=4\pi G(\rho+p)
\end{equation}
By applying $r=1/H$ and $\dot{r}=-\dot{H}/H^{2}$,we get
\begin{equation} \label{p}
\dot{H}=-4\pi G(\rho+p)
\end{equation}
That equation (\ref{p}) is rewritten by adding the definitions of densities and pressures as
\begin{equation} \label{q}
f_{A}\dot{H}=-\frac{\kappa^{2}}{2f_{\mathcal{T}}}\left(\rho_{m}+p_{m}+\rho_{\mathcal{T}}+p_{\mathcal{T}}\right)
\end{equation}
After applying to the conservation equation and conducting a temporal integration we have
\begin{equation}\label{r}
\frac{2}{2-\Delta}f_{A}H^{2}=\frac{\kappa^{2}}{3f_{\mathcal{T}}}\left(\rho_{m}+\rho_\mathcal{T}\right)
\end{equation}
Note that considering $\Delta=0$ or $f(\mathcal{T})=\mathcal{T}$, respectively, Eqs. (\ref{q}) and (\ref{r}) are reduced to the standard form for Barrow and $f(\mathcal{T})$ separately.

It is possible to do another analysis. The non-conservation relation is utilized, if the non-equilibrium case is preferred, i.e.
 \begin{equation}
\dot\rho_{\mathcal{T}}+3H\left(\rho_{\mathcal{T}}+p_{\mathcal{T}}\right)=\frac{3}{\kappa^{2}}\frac{2}{2-\Delta}f_{A}H^{2}\dot{\mathcal{T}}f_{\mathcal{TT}}
\end{equation}
Furthermore, the entropy correction is taken as
\begin{equation}
S=\frac{f(A)}{4G}f_{\mathcal{T}}
\end{equation}
The $f(\mathcal{T})$ correction acts on the physics of the system, specifically as a correction of the gravitational constant, as $G_{eff}=G/f_{\mathcal{T}}$. In contrast, the Barrow correction improves how the horizon area is calculated \cite{jb}. This leads to the definition of the new entropy. Moreover, this entropy form aligns with the dark energy component proposal to yield a thermodynamic relation that recaptures all the features of the Barrow and $f(\mathcal{T})$ cases separately. The thermodynamic relationship with these ingredients takes the form
\begin{equation}
Tds=dE-WdV+T\left(\frac{4-\dot{r}+\Delta\dot{r}/2}{2-\dot{r}-\Delta(1-\dot{r}/2)}\right)\frac{f(A)}{4G}df_\mathcal{T}
\end{equation}
The final term may originate from an out-of-equilibrium system, as was previously discussed \cite{ys,tn}.
As an example, the Barrow $f(\mathcal{T})$ field equation will be solved and the state parameter in this particular context will be examined in the following section.
\section{Application on the generalized new type of $f(\mathcal{T})$ model}
\begin{figure}[hbt!]
\begin{center}
\includegraphics[height=2.5in]{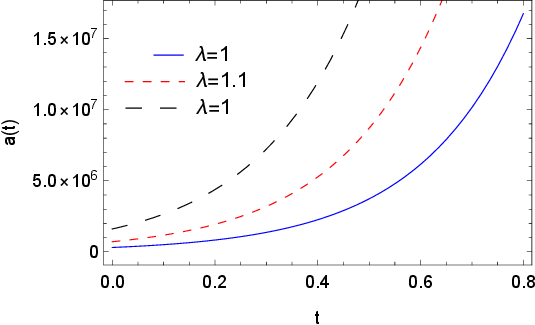}
\caption{Plot of the scale factor $a(t)$ against time $t$ for different values of the model parameter $\lambda$. The other parameters are considered as $G=1$,~ $\kappa=1$,~$\rho_{m0}=1$,~ $C_{1}=1$.}
\label{figscale}
\end{center}
\end{figure}

To see how a $f(\mathcal{T})$ model may be affected by the addition of Barrow entropy, we use a toy model of $f(\mathcal{T})$ gravity and explore the cosmological evolution. This model is taken as,
\begin{equation}
f(\mathcal{T})=\lambda \mathcal{T}^{2}
\end{equation}
where $\lambda$ is a model parameter taking constant values. We use the above $f(\mathcal{T})$ model in Eq.(\ref{r}) and obtain an approximate solution of the scale factor for $\Delta=0.01$ as,
\begin{equation}
a(t)\approx\frac{1}{576\pi^{2}\lambda^{2}}\left[e^{-2\sqrt{2\kappa\pi/6}~t-12\sqrt{\pi\kappa\lambda}~C_{1}}\left(e^{2\sqrt{2\kappa\pi/6}~t+12\sqrt{\pi\kappa\lambda}~C_{1}}+G\pi\kappa\lambda\rho_{m0}\right)^{2}\right]
\end{equation}
where $C_{1}$ is the constant of integration. Since the model has become quite complex after the introduction of Barrow entropy in $f(\mathcal{T})$ gravity, we faced many challenges while finding the exact solution of the field equations to find the expression for the cosmological scale factor. So we found the above approximate solution by fixing the Barrow parameter to a particular value. But the above solution of the scale factor preserves most of the flavor of the modification introduced by the Barrow entropy in the $f(\mathcal{T})$ gravity. To check the evolution, we generate a plot of the scale factor $a(t)$ against time $t$ in Fig.(\ref{figscale}). We see that the scale factor increases exponentially with the evolution of time, which is expected for the $f(\mathcal{T})$ model considered. From the figure, we also see that there is a prominent dependency of the scale factor on the parameter $\lambda$ which becomes stronger as time evolves. This is evident from the fact that the different trajectories move farther apart from each other as time evolves.

So the solution is acceptable and we proceed to explore the evolution of the universe described by this model. The equation of state (EoS) parameter is given by,
\begin{equation}
\omega=\frac{p_{m}+p_{de}}{\rho_{m}+\rho_{de}}
\end{equation}

\begin{figure}[hbt!]
\begin{center}
\includegraphics[height=2.5in]{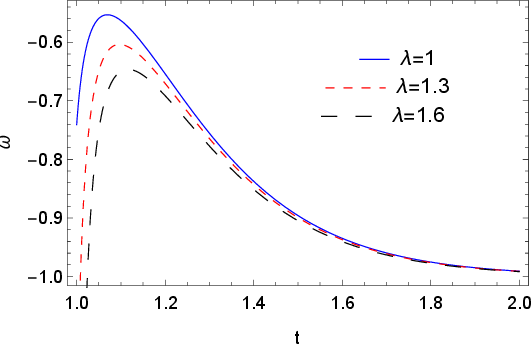}
\caption{Plot of the equation of state parameter $\omega$ against time $t$ for different values of the model parameter $\lambda$. The other parameters are considered as $\Delta=0.01$, $G=1$,~ $\kappa=1$,~$\rho_{m0}=210$,~ $C_{1}=0.1$.}
\label{figeos}
\end{center}
\end{figure}

\begin{figure}[hbt!]
\begin{center}
\includegraphics[height=2.5in]{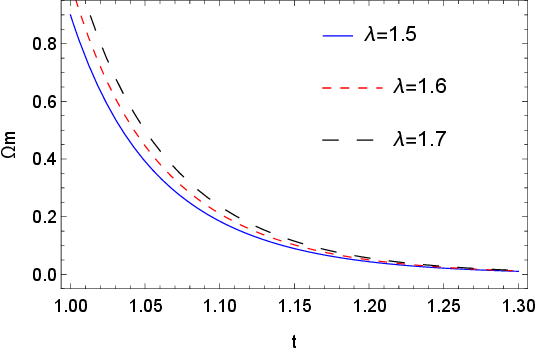}
\caption{Plot of the dimensionless matter density parameter $\Omega_{m}$ against time $t$ for different values of the model parameter $\lambda$. The other parameters are considered as $G=1$,~ $\kappa=1$,~$\rho_{m0}=210$,~ $C_{1}=0.1$.}
\label{figomegam}
\end{center}
\end{figure}

\begin{figure}[hbt!]
\begin{center}
\includegraphics[height=2.5in]{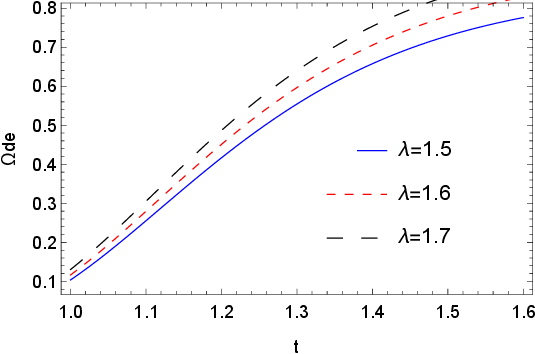}
\caption{Plot of the dimensionless dark energy density parameter $\Omega_{de}$ against time $t$ for different values of the model parameter $\lambda$. The other parameters are considered as $\Delta=0.01$, $G=1$,~ $\kappa=1$,~$\rho_{m0}=210$,~ $C_{1}=0.1$.}
\label{figomegade}
\end{center}
\end{figure}

\begin{figure}[hbt!]
\begin{center}
\includegraphics[height=2.5in]{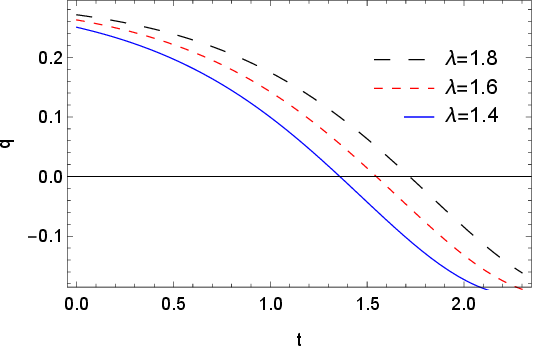}
\caption{Plot of the deceleration parameter $q$ against time $t$ for different values of the model parameter $\lambda$. The other parameters are considered as $G=1$,~ $\kappa=1$,~$\rho_{m0}=8$,~ $C_{1}=-0.3$.}
\label{figdec}
\end{center}
\end{figure}

The EoS parameter $\omega$ is plotted against time $t$ in Fig.(\ref{figeos}). From the figure, we see that the EoS parameter is in the range of $-0.5$ to $-0.9$, eventually settling around $-1$ as time evolves. So we have $\omega<-1/3$ showing a quintessence regime before settling down around the $\Lambda$CDM regime with the passage of time. The figure also shows that in the early universe, there is a clear dependency of $\omega$ on $\lambda$, but as time evolves the dependency on $\lambda$ decreases. It shows the clear tendency of the EoS parameter to settle down near $-1$ irrespective of the parameter space. The dimensionless density parameters are given by,
\begin{equation}
\sigma_{m}=\frac{8\pi G}{3H^{2}}\rho_{m} 
~~~ \text{and} 
~~~~\sigma_{de}=\frac{8\pi G}{3H^{2}}\rho_{de} 
\end{equation}
The plots for the dimensionless density parameters are obtained in Figs.(\ref{figomegam}) and (\ref{figomegade}). From Fig.(\ref{figomegam}) we see that there is a decay in the matter density of the universe as time evolves. From the figure, we see that there is a prominent dependency on $\lambda$ in the early universe but it wanes away at late times. From Fig.(\ref{figomegade}) we see that there is a growth in the dark energy density with the evolution of the universe. This is perfectly consistent with the picture of an energy-dominated universe leading to an accelerated expansion scenario. From the figure, we also see that initially the trajectories almost coincide showing very little dependency on $\lambda$. But in the late universe, this dependency becomes more prominent. In Fig.(\ref{figdec}) we have plotted the deceleration parameter against time $t$ for different values of the model parameter $\lambda$. It is seen that initially, $q$ is at the positive level indicating a decelerating expansion in the early universe. But as the universe evolves the value of $q$ decreases, and finally there is a clear transition from a positive level to a negative level. This shows that the late universe enters an accelerated expansion scenario, which perfectly matches with the observations. It is also understood from the plot that as $\lambda$ increases the transition from decelerated to the accelerated universe is more delayed. Note that in the above we have not included the expressions for $\omega$, $\Omega_{m}$, $\Omega_{de}$, and $q$ because the expressions for these parameters are complicated and voluminous.

\section{Conclusion}
Black hole thermodynamics provides a means of applying the laws of thermodynamics to a new approach to studying the gravitational field equations, building upon the work of Hawking and Bekenstein. This paper considers Barrow entropy and derives the Friedmann equation using the gravity-thermodynamics conjecture. The possibility that quantum-gravitational effects could alter a black hole's actual horizon area is taken into consideration in Barrow's model. Stated differently, this model increases the black hole area to create a fractal horizon surface. The first law of thermodynamics in this gravitational model is examined, and $f(\mathcal{T})$ gravity is taken into consideration using the concepts of the gravity-thermodynamics hypothesis. The primary research goal of this work was to integrate $f(\mathcal{T})$ gravity and Barrow entropy under one framework to analyze the universe's dynamic evolution over time. The torsional effect of the $f(\mathcal{T})$ gravity and the fractal nature of Barrow entropy impose serious corrections in the dark energy model. Our findings demonstrate that for both the individual cases and the union of the Barrow and $f(\mathcal{T})$ models, the scale factor and the state parameter behave as predicted. We have also studied the trend of the density parameters and the deceleration parameter. The results are in compliance with the observations. The findings also indicate that the Barrow parameter $\Delta$ has a considerable impact on the new model of dark energy. Future research will examine the limitations of its values in this novel setting. A reconstruction scheme can be set up with this new model to derive $f(\mathcal{T})$ gravity models with a Barrow entropy effect embedded in it. A dynamical system analysis and a perturbation analysis of the model will be really interesting. All these can be very good future projects related to the newly developed model. This form of dual corrections to the standard model is a really interesting way to create new models that may open up new avenues in cosmology.

\section*{Acknowledgments}

P.R. acknowledges the Inter-University Centre for Astronomy and Astrophysics (IUCAA), Pune, India for granting visiting associateship.

\section*{Data Availability}

No new data were generated in this paper.

\section*{Conflict of Interests}

There are no conflicts of interest in this paper.

\section*{Funding Statement}

No funding was received for this paper.


\end{document}